\documentclass[prb,reprint,superscriptaddress]{revtex4-1} 

\usepackage{amsmath}
\usepackage{amsfonts}
\usepackage{graphicx}
\usepackage{soul}
\usepackage{color}

\begin{document}

\title{Working with writing experts to enhance a physics curriculum}

\affiliation{The George Washington University, Department of Physics, 725 21st Street NW, Washington, DC 20052, USA}
\affiliation{The George Washington University, University Writing Program, 2100 Foxhall Road NW, Washington, DC 20007, USA}
\affiliation{The George Washington University, Women's, Gender \& Sexuality Studies, 801 22nd Street NW, Washington, DC 20052, USA}
\affiliation{American Association of Physics Teachers, 1 Physics Ellipse, College Park, MD 20740, USA}


\author{Alexander J. van der Horst}
\email{ajvanderhorst@gwu.edu}
\affiliation{The George Washington University, Department of Physics, 725 21st Street NW, Washington, DC 20052, USA}

\author{Rachel C. Riedner}
\affiliation{The George Washington University, University Writing Program, 2100 Foxhall Road NW, Washington, DC 20007, USA}
\affiliation{The George Washington University, Women's, Gender \& Sexuality Studies, 801 22nd Street NW, Washington, DC 20052, USA}

\author{Gary D. White}
\affiliation{The George Washington University, Department of Physics, 725 21st Street NW, Washington, DC 20052, USA}
\affiliation{American Association of Physics Teachers, 1 Physics Ellipse, College Park, MD 20740, USA}

\author{Carol L. Hayes}
\affiliation{The George Washington University, University Writing Program, 2100 Foxhall Road NW, Washington, DC 20007, USA}

\author{Axel W. Schmidt}
\affiliation{The George Washington University, Department of Physics, 725 21st Street NW, Washington, DC 20052, USA}

\author{Evangeline J. Downie}
\affiliation{The George Washington University, Department of Physics, 725 21st Street NW, Washington, DC 20052, USA}

\author{William J. Briscoe}
\affiliation{The George Washington University, Department of Physics, 725 21st Street NW, Washington, DC 20052, USA}

\date{\today}

\begin{abstract}
A common observation of physics faculty and professionals hiring physics graduates is that undergraduate students have deficient physics-specific writing skills, in particular difficulties with writing genres such as abstracts, reports, articles, proposals, and posters. Students are not often exposed to such authentic writing experiences as part of their curriculum, and faculty may not feel qualified to teach writing. Scholarship focused on writing transfer, defined as the ability to apply the writing skills and knowledge learned in one context in a new writing context, suggests that students need support with transfer-focused writing skills such as genre and audience awareness and that these are skills in which students can be trained. For physics students, physics faculty are best placed to do this as successful participants in their disciplines. We present an interdisciplinary effort, by faculty of physics and writing studies, of developing a writing in physics curriculum using a scaffolded approach including three courses in the second through fourth year of an undergraduate physics program. This approach uses threshold concepts from writing studies, such as genre, audience, and purpose. Students are given the methodological tools to approach writing, and the instruction includes structured peer feedback that is connected to the instructor's feedback and grading. This effort led to improvements in the written and oral communication skills of students in a broad range of genres and for various audiences, evidenced, for instance, by an increase in successful fellowship applications. The course materials developed for this effort are publicly available to the community.
\end{abstract}

\maketitle

\section{Introduction}

Communicating scientific progress to a wide range of audiences has been a core activity of physicists for centuries. However, acquiring the skills necessary to communicate effectively has not been at the core of the physics curriculum \citep{Lewis2025}. When it comes to written communication, there is often no formal training, and students are expected to develop these skills by osmosis. Many physics faculty recognize that students lack the physics-specific writing skills necessary for professional writing practices, but they generally do not have experience and expertise in teaching these writing skills \citep{Reynolds2012}. This recognition by faculty is not just anecdotal: a report by the Joint Task Force on Undergraduate Physics Programs (J-TUPP), for example, highlighted the importance of written communication and recommended that more attention be paid to developing this skill in the undergraduate physics curriculum \citep{JTUPP2016}. It is also important to note that writing can improve conceptual understanding and support the research process \citep{Hand2001,Reynolds2012,AdlerKassner2022}. Here we present an interdisciplinary effort between faculty from the Department of Physics and the University Writing Program at a research-focused university, resulting in revisions within its undergraduate physics program. The revisions include a focused effort to develop written and oral communication skills using writing pedagogy through three core courses spread over the second through fourth years of a four-year undergraduate program. 

Writing skills are key for physics majors but can be challenging to attain. Although physicists study the world around us, the terminology used in physics is abstract and is based on mathematical terms. To succeed academically, students must show that they have learned to communicate as professional physicists in these abstract terms. To succeed outside of the university context, students must be able to communicate with future employers, proposal reviewers, and non-expert audiences who may not understand technical physics language. In addition, for most course assignments, students turn in work, e.g., a problem set, to a professor who already knows the correct answer. This is rarely the case in real world situations faced by physicists. When expert physicists write, it is often for an audience that lacks foreknowledge of the answer to the problem at hand, and success requires the construction of an effective argument that is accessible to the audience.

For these reasons, in addition to the training in physics content, students need to develop transfer-focused writing skills, i.e., the ability to transfer written communication skills learned in one context and adapt these skills to new audiences and contexts \citep{Moore2017}. Common types of writing are governed by different conventions of syntax, format, and style, and they serve a certain rhetorical purpose, targeted for a particular audience. Scholarship focused on writing transfer suggests that students need support with transfer-focused writing skills such as genre and audience awareness, and that these are skills in which students can be explicitly trained \citep{Bergmann2007,Beaufort2007,Nowacek2011,Elliott2016,Driscoll2020}. This instruction can be provided by physics faculty members who are successful participants in their disciplines \citep{Thompson2021}. 

Genres are typical forms of written communication that have specific, recognizable, and teachable expressions within disciplines, and are critical to learning, participation, and success within a discipline \citep{Miller1984,Swales2004,AdlerKassner2015}. In physics, typical genres include abstracts, posters, proposals, articles, and more. Such genres provide the written means to achieve specific goals \citep{Miller1984}, such as securing funding or sharing knowledge with colleagues. Genre is a threshold concept, a core disciplinary capacity that enables students to progress in their studies and transition to professional settings \citep{Meyer2003,AdlerKassner2015}. The curricular intervention presented here builds on the recognition that genre instruction is a key competency of undergraduate education and specifically for disciplines such as physics. However, physics faculty do not necessarily have the training to teach genres effectively. Understanding threshold concepts such as genre, audience, and purpose is a complex teaching effort. The curriculum presented here, together with publicly available teaching materials \citep{CapstoneGithub}, is meant to support physics faculty in incorporating writing instruction into their courses and curriculum.

In this paper, we discuss a framework in which research-based writing pedagogy is applied in a physics context, and curricular elements are intentionally connected across multiple semesters in a scaffolded approach. In Section~\ref{sec:motivation}, we provide motivation for a writing curriculum in physics and discuss the development areas for both students and faculty. In Section~\ref{sec:innovations}, we present the curricular innovations for writing in physics that were implemented. The scaffolding of the program is discussed in Section~\ref{sec:scaffolding}, and in Section~\ref{sec:materials}, we discuss the transferability of course materials and activities. We summarize and conclude in Section~\ref{sec:conclusions}.

\section{Motivation for a Structured Writing in Physics Curriculum}
\label{sec:motivation}

Writing is taught sporadically in the undergraduate physics curriculum \citep{Lewis2025}. There are various reasons put forward by the faculty for why this is the case. For example, many faculty feel that explicitly dedicating class time to developing skills such as written and oral communication would lead to sacrificing the teaching of content knowledge in their field \citep{Carter2007}. Some faculty might presume that these skills can be acquired by students without formal training, and few physics faculty received physics-specific writing instruction themselves in their own education \citep{Lewis2025}. At the same time, faculty would like their students to succeed in a career after graduation. The J-TUPP report \citep{JTUPP2016} emphasized that the physics curriculum must prepare students for a variety of careers and suggested revisions of the curriculum with an emphasis on skills such as being able to work well in teams and employing computational tools, as well as writing and speaking for a variety of audiences. These skills are useful not only for students pursuing careers outside of academia, but also for students continuing in research or becoming faculty, who spend a significant amount of their time writing in various genres and presenting for different audiences.

To succeed in their careers, physics students often need to be able to use science in non-academic settings, and to transfer their insights by writing well for various audiences. They often have to demonstrate their ability to communicate with other scientists in their field or with general science professionals, but also with funding agencies, potential employers, and public audiences \citep{Thompson2021}. To achieve this goal, students must understand the genre, purpose, and audience of their writing \citep{Moskovitz2011}. In the learning process, students will have to move from a writer-centered mindset, focused on what information they need to put on paper, to an audience-centered mindset, focused on who the audience is, what they need to understand, and how writing can help them get to that understanding. This often includes organizing information into the form of an argument, and students will have to master genre-specific technical elements, such as writing around figures, tables, and equations.

Mastering physics genres, the different forms of writing that physicists use to communicate with each other and with external audiences, is a means through which students signal disciplinary knowledge and expertise \citep{AdlerKassner2015}. There are crucial differences between an abstract for a presentation at a specialized academic conference, an abstract for a publication in a peer-reviewed journal, and an abstract for a proposal requesting research support. Although these are all written abstracts, their purpose, audience, and style are markedly different. Mastering these genres is crucial to signaling disciplinary expertise and belonging to the physics community, and understanding and communicating through physics genres is a necessary educational outcome. However, faculty training typically does not include preparation for teaching writing, science communication in general, and concepts developed in writing studies such as genre and audience.

Physics faculty may not feel qualified to teach writing because they associate teaching writing with grammar instruction \citep{Neff2009}. The first step in changing how writing is conceived and taught is that faculty need to recognize that successful writing is not defined by grammar alone and that writing practices are not generic. What constitutes successful writing is different between disciplines and has been shaped by the types of knowledge produced in those fields, as well as by the needs of the audiences addressed by members of those disciplines \citep{Carter2007}. Faculty are the ones best positioned to teach writing within their discipline because they are the ones with expertise in writing to the specific audiences targeted by their field, in genres that serve the purposes of their field. For example, physics faculty know what a physics abstract, research proposal, or poster should look like, including how each of those genres might differ when written for different audiences. Therefore, they are best positioned to provide instruction and feedback to students on how to write in those genres, such as what types of information to include, in what order, and with what level of jargon for the anticipated audience and purpose. It is also important to realize that writing requires sustained instruction over multiple semesters to accomplish increasingly sophisticated learning as students progress through their studies and into professional careers.

When students are taught disciplinary genres, they learn to communicate through the accepted written practices of a disciplinary community. This communicative competency signals to disciplinary professionals that they are members of the community \citep{Florence2004,Fankhauser2025}. When students are recognized as participants, they become disciplinary and professional participants and start to see themselves as such, thus creating identity as members of the community \citep{AdlerKassner2015,Francis2022}. As a result, these types of curricular interventions set students up for success, not only within the curriculum but also in professional settings. 

The J-TUPP report \citep{JTUPP2016} provided guidance for the revision of an undergraduate curriculum and stated that broad learning goals, such as those associated with communication skills, can be incorporated into the curriculum at multiple points or can be coherently addressed in a single integrative experience such as a course or a sequence of courses. Having similar skill development activities at multiple points in a curriculum will enhance and deepen the development of those skills, so having a comprehensive written communication effort at various points in a curriculum will lead to better development of writing skills \citep{Lang2016}. Here, we show an approach and implementation that addresses curricular innovations for writing in physics based on an interdisciplinary collaboration between physics and writing faculty.

\section{Curricular Innovations for Writing in Physics}
\label{sec:innovations}

The faculty of our Department of Physics has revised the curriculum of our undergraduate Bachelor of Science programs, both on a course-by-course basis and at an overall program level. This has been done by creating or revising various courses and infusing existing courses with relevant activities. The courses involved in the overhaul of the department’s instructional efforts and related to written communication are a 3-credit Intermediate Laboratory course that is taken in the second or third year; a 3-credit Physics Capstone course that is taken in the third year; and a 1-credit Physics Symposium course that is taken in the fourth year. These three courses are required for all physics majors, and the first two courses are approved as Writing in the Discipline (WID) courses by the University Writing Program. This label certifies, for example, that at least 40\% of the grade is determined from discipline-appropriate writing assignments and that assignments involve faculty and peer feedback, revision, appropriate scaffolding, and attention to genre and audience. The university requires students to take a University Writing course in their first year, followed by at least two WID courses in the following years. 

Having two physics courses with the WID designation allows physics majors to achieve their WID graduation requirement within their major. These courses are embedded in a curriculum typical of most physics degree programs in the country \citep{SPINUP2003}. The intentionality of expanding and connecting the writing and communication experiences across semesters in the three aforementioned courses is what we consider a particularly notable aspect of the curricular innovations. Here, we describe the general approach and implementation in the three courses in more detail. In addition to the various writing activities inside and outside the classroom, we also discuss the importance of peer feedback and how it is incorporated into these courses. This is a crucial component of writing pedagogy and has been an important and effective tool in revising the way in which we teach physics writing. We emphasize that the writing instruction in all of these courses is provided by physics faculty.

\subsection{Pedagogical Approach}
\label{sec:approach}

The primary aim of our approach is to enable students to communicate information successfully with both physics-focused and non-technical audiences. Our approach centers on giving students practice with more authentic writing in the physics discipline, favoring forms such as abstracts, proposals, manuscripts, semester-long lab research notebooks, letters to colleagues, and posters over activities characteristic of school work, such as one-off written lab reports and problem sets that are seen only by the instructor. This focus on authentic physics writing is done through the development of a sequence of connected courses in which genres are explicitly taught. Genres are forms of writing whose structures have been shaped to serve specific purposes for members of their community \citep{Miller1984}. In writing for those specific audiences and purposes, genres develop typical features. Writing studies scholars see the definition of ``good writing'' as discipline specific \citep{Kruse2012}: each discipline has its own written genres with different expectations regarding what constitutes evidence, formatting, modes of address, syntactical style, etc. Developing genre awareness can play a key role in helping students write successfully as they move into new contexts. Physics students must learn to write in physics contexts and may be more successful at these writing tasks if they are encouraged to focus on the specific audience and purpose while writing.

In developing the new physics writing curriculum, we incorporated pedagogical strategies that taught students explicitly how to write in new genres via the CAT (Collect, Analyze, Try) approach. This approach, developed by the Writing Transfer Project research team \citep{Driscoll2020}, first asks students to collect multiple samples of the new genre in which they are writing. In our physics writing curriculum, instructors sometimes provide those samples; at other times, students are asked to find their own. Once students have collected samples, they are prompted to analyze those genre samples for similarities and differences in relation to each document's target audience and purpose, structural features, organization, sentence length, and use of jargon. If all of the samples share certain features, those are features that the student will want to imitate. If certain features only appear in one or two samples, students can decide if including them is appropriate to the target audience and the purpose of their writing. The final stage of the CAT approach is for students to try writing in the genre themselves.

The physics writing curriculum was designed to offer multiple opportunities over time to engage in scaffolded writing practice in new genres. The cognitive demands of writing are intense and take time, and this cognitive load can be reduced through deliberate practice of component processes \citep{Kellogg2009}. The repeated teaching of the CAT approach, sustained writing practice, and support provided by the three-course sequence mean that students asked to write in a new genre are less likely to be overwhelmed with simultaneously having to figure out the content they want to convey and the conventions of a new genre. Having an already established, familiar process for analyzing genre can help students engage in writing with more attention available for the coordination and control of both genre elements and content \citep{Kellogg2009}, because the work on the component processes has become familiar. Students can draw on the CAT approach when encountering new genres in any context, from writing in physics disciplines to other disciplines, to job applications.

To emphasize the importance of genre as a threshold concept that can be learned over time with support, our physics writing curriculum includes structured peer feedback \citep{Thompson2021}, where students use rubrics and prompts designed by instructors to emphasize key evaluative criteria for that type of genre. This peer feedback serves multiple purposes. First, it may help acculturate students into scientific professional communities by giving them experience with this important form of interaction between scientists \citep{Guilford2002,Trautmann2009}. Second, it may help students better understand that they are writing to an audience via genres of the discipline, especially if students are asked to take on specific professional roles as they are giving their colleagues feedback, e.g., giving feedback from the perspective of a proposal reviewer, a physicist from a different subdiscipline, or a general audience member. Third, in our approach, peer feedback rubrics are often aligned with the grading rubrics used by instructors, adding another layer of purpose to peer feedback activities.

\subsection{Intermediate Laboratory Course}
\label{sec:labcourse}

The Intermediate Laboratory course is typically taken by students in the spring semester of their second or third year and is considered a bridge course from the introductory physics courses to the upper-level curriculum. Historically, the course consisted of revisiting key physics experiments from the late 19th and early 20th centuries, such as the cathode ray beam, circuit analysis, hydrogen spectrum, speed of light, and single and double slit diffraction. Students wrote one-off lab reports that describe the nature of the experiments and provide ``accepted'' answers and explanations related to the experiments. There was no mention of genre, virtually nothing related to audience or purpose, and hardly any peer feedback or iteration of drafts, which left little opportunity for students to improve their writing skills.

This course has evolved considerably over time, and the teaching of the instructor has gradually been influenced by the input of writing faculty. As the instructor worked to change the course, the writing components of the course were modified such that the assignments related to more authentic forms of physics writing. Instead of relying on one-off lab reports, only seen by the instructor with no peer feedback or iterative improvements to learn from, the updated course required students to keep a lab notebook, documenting what they were exploring each week. This was modeled after lab notebooks that scientists in industrial research laboratories keep and was based on a model gleaned from a similar course taught by Prof. T. Gfroerer at Davidson College. Various genres of physics writing, with explicit emphasis on audience and purpose, were introduced via the CAT method, including abstracts, technical letters to colleagues, research posters, and slides for presentations.

Scaffolding of written and oral communication activities is now part of this course. A sequence of assignments might include investigating the behavior of a cathode ray beam under various conditions, documenting details in the student’s research notebook over several weeks of the course, writing an abstract appropriate for a journal, and then giving an oral presentation appropriate for an audience of the student's peers, synthesizing what has been learned about the properties of the beam and how it relates to more modern ideas about electrons. The various assignments include opportunities for peer and instructor feedback, as well as feedback-based revisions. 

It should be noted that the instructor reports that their early attempts at teaching genre within the course were largely ineffective until they learned to incorporate various intentional and specific practices such as the CAT method, providing experiences with authentic genres within the discipline, and guided peer feedback with rubrics. Clear progress can be seen in the multiple drafts of student writing and in the increase in high-quality student writing at various research presentation events.

\subsection{Physics Capstone \& Physics Symposium Courses}
\label{sec:capstone}

Within the program discussed here, Bachelor of Science physics students participate in at least one large research project, in which they are supervised by a faculty member and sometimes co-supervised by one of their close collaborators or a graduate student. In addition to the large research project, which should start in the third year and be finalized in the fourth year, students are encouraged to do other research projects within the department. Students are also encouraged and supported by the physics faculty and the Center for Career Services of the university to apply for research experiences and internships outside the department during the summer months. To support them through their research experiences, every Bachelor of Science physics student takes the Physics Capstone and Physics Symposium courses. 

The original Physics Capstone course, before the curricular redesign, was taken by students in the fall semester of their fourth year, loosely structured with poorly defined deliverables, and lacking in effectiveness. In the revised course sequence, the Physics Capstone course was moved to the fall of the third year and restructured to include instruction in, and production of, various writing genres. Furthermore, a new course was added in the spring semester of the fourth year, the Physics Symposium course, to give students the support and guidance they need for the research experience during their final semester at the university. In the new structure, communication activities evolve from short writing pieces and presentations, such as abstracts and research elevator pitches, to longer and higher-stakes activities, such as papers and conference-style posters and presentations. The courses now better prepare students to make decisions regarding what to do after graduation by including discussions of both academic and non-academic career paths, and they include guest lectures by staff of the Center for Career Services, as well as other experts on careers inside and outside academia.  

Both the Physics Capstone and Physics Symposium courses are taught in a highly collaborative mode, consistent with modern, discipline-based, education-research-supported pedagogy, with minimal lecturing and maximal active student participation. There are a variety of activities, including discussions on topics such as research methodology and ethics, interactions of physicists with society, and issues related to fostering a diverse physics community. The development of written and oral communication skills occurs in tandem through different topics. The CAT method is used as a basis for writing instruction, and there are explicit discussions on giving presentations, including preparation, purpose, style, and the content and structure of slides. Students first write an abstract for a Physics Today article of their choosing and give a short presentation without slides on this article; they lead an in-class discussion and give a presentation with slides on a research article in their area of interest; and they write an abstract and give a presentation on the research project that they will carry out in their third and fourth year. Students write a resume and cover letter for a specific internship or job opportunity of their choice and give an elevator pitch in class. They also write an abstract for a proposal related to their research or interests, and they give a presentation on a proposal for an outreach activity that could be carried out by our physics student organization. The sequence of assignments deliberately models and teaches authentic physics genres, and it does so in ways that give students multiple opportunities to practice engaging with those genres. 

All these written and oral communication activities include peer feedback in various ways. Toward the beginning of the semester, students are asked to comment on genre expectations, e.g., if an abstract written by one of their peers matches the audience, length, organization, and format of the genre. They are also asked to summarize the main point of the article in one sentence and share this with the author, leading to discussions on how a piece of writing can be interpreted differently by different readers. In other peer feedback activities, students are asked to identify and discuss with their peers what is most important, interesting, or striking about a piece of writing; and also what is almost but not quite said \citep{Elbow1989}. In some of these activities, students are also asked to constructively discuss strengths and weaknesses based on a rubric that matches the instructor's grading rubric. Students are not giving each other grades, but they are discussing with each other the various rubric items, such as content, motivation, organization, referencing, vocabulary, voice, and fluency. A peer feedback activity that is most liked by the students because it models authentic disciplinary practice is a review panel for proposal abstracts with students serving as the reviewers, structured in a way that is similar to peer review panels that professional physicists participate in.

The main learning goal of the Physics Symposium course is for students to communicate physics research in a manner that is appropriate for the discipline and various audiences. They write a final report on the research they perform in between the Physics Capstone course and this course, in an appropriate format and style; present a poster on their research at a college-wide Research Showcase; and give an oral presentation on their research for faculty and students at the Physics Undergraduate Symposium Day. Feedback is provided by the instructor and the student's peers in this course as well. For the final report, the students write the introduction and conclusion sections for a general physics audience, so their work is accessible to a broader audience, and their peers can provide meaningful feedback. The other sections of the report typically use more jargon and both research advisors and the course instructor provide feedback. The audience for the college-wide Research Showcase poster is much broader, since the college includes departments in the natural and social sciences, humanities, and arts, and students will potentially present their poster to students and faculty across all those disciplines. The culmination of written and oral communication skills will prepare students well for many different careers, both inside and outside of an academic and/or research environment.

\section{Scaffolded Curriculum \& Outcomes}
\label{sec:scaffolding}

It is important to emphasize that there has been coordination between the different physics instructors, with support from the writing faculty, to ensure scaffolding in the curriculum. The first example of this is the CAT approach, which is introduced and discussed early on in both the Intermediate Laboratory and Physics Capstone courses. Students typically take the Intermediate Laboratory course in the spring semester of their second or third year and the Physics Capstone course in the fall of their third year (and only in exceptional cases the latter is taken in the fall of their fourth year). This repetition solidifies the CAT approach as a methodology that they will use in their writing activities and assignments whenever they are faced with a new type of writing.

A second example of scaffolding are certain genres that are common between the different courses. Students are required to develop and present a scientific poster on a research project they conducted in the last month of the Intermediate Laboratory course. As part of the Physics Symposium course, they develop and present a scientific poster on their large research project that they perform in their third and fourth year. The audience of these posters is different: for the Intermediate Laboratory course, the students present their posters to other physics students and faculty at an end-of-semester event, whereas the Physics Symposium posters are presented for a much broader audience at a college-wide Research Showcase. Doing this in a guided manner prepares students well for presenting posters at other venues, which some of them have done at conferences of the American Physical Society and the American Astronomical Society. Furthermore, while presenting a poster at the college-wide Research Showcase is now required for Bachelor of Science physics students in their final year, this was not the case early on in the implementation of this physics writing curriculum. The growing emphasis on communication in the curriculum led to a strong increase in the number of fourth-year physics students presenting at this venue; and now several third-year students who were in the Intermediate Laboratory and Physics Capstone courses also present posters, something that was rare before these curricular changes.

Another important example of scaffolding in the curriculum is the way in which peer feedback is implemented. Although there is a variety of different activities included in all three courses, there are also some similar activities during the semesters and between courses to solidify the practice of peer feedback for different contexts. The peer feedback activities for some of the abstracts that students write in the Physics Capstone course are very similar, and also similar to the peer feedback activities for when they write the introduction for their final paper in the Physics Symposium course. These activities include, for instance, identifying and discussing what is most important, interesting, or striking, and what is almost but not quite said. The result of repeating similar feedback activities is that once students are providing peer feedback on other sections of the final report, they do not need prompting to automatically carry out certain feedback activities, showing that they have internalized this as a skill they can use in the future.

Another example of the outcome and impact of the physics writing curriculum in a scaffolded approach is that our physics students are applying for more competitive research opportunities and are more successful in obtaining funding to support their research. They are accepted into more external research opportunities, but the most striking and tangible is the success rate of obtaining fellowships to carry out research supervised by a faculty member at our university. These fellowship applications often require students to motivate their project in writing for a non-technical audience, a genre that our students previously struggled with. For many years, there was on average one physics student per year obtaining a college-wide or university-wide research fellowship, but in the last few years this number has increased very significantly, with in the last few years on average four of such fellowships being awarded to physics students, which is 20$-$25\% of the total number of awards. This is particularly noteworthy because the Physics Department is one of more than 40 departments in a college of arts, humanities, social sciences, and natural sciences. In our view, this can be attributed (at least in part) to the focus on threshold writing concepts such as genre, purpose, and audience in these physics courses, which prepares students well for applying to these opportunities.

\section{Transferable Curricular Materials}
\label{sec:materials}

A challenge in teaching the courses described here, in particular the written and oral communication activities, is that there are no standard textbooks for them. Resources are available, but they are scattered on websites from many institutions. Therefore, all course materials for the Physics Capstone and Physics Symposium courses, including syllabi, slides, rubrics, etc., as well as the writing activities in the Intermediate Laboratory course, are available online \citep{CapstoneGithub}. This effort was part of the American Physical Society's PIPELINE Network project, which aimed at developing new approaches to teaching physics innovation and entrepreneurship. We were one of the six partner universities, focusing its efforts on written and oral communication through the aforementioned courses, and many of the materials were developed in our interdisciplinary collaboration.

The writing elements of the courses ensure that the students develop genre and audience awareness. Students are also taught the habits and strategies for producing writing through peer feedback and guidance with revision and editing. The oral communication aspects of the courses have been developed following a similar pedagogy and with overlapping materials, making a clear connection between written and oral communication.

For assessing the various learning objectives of the three physics courses for genre instruction, rubrics for evaluation of the students' products have been developed. For the final presentation of the Physics Capstone and Physics Symposium courses, students' performance is assessed on the basis of content, motivation, subject knowledge, organization, communication aids, elocution, body language, eye contact, and enthusiasm. The final research paper has similar assessment categories, with some oral presentation competencies replaced by written competencies such as genre, referencing, vocabulary, voice, fluency, and conventions. For resumes and cover letters, students are assessed in categories such as genre, target audience, structure, tone, and presentation. These same rubrics are part of the peer review process in the courses, so that students have complete insight into the evaluations by the instructor. These rubrics are part of the course assessment as required by the university, and some are also part of the overall assessment of the undergraduate programs, since the Physics Capstone and Physics Symposium courses provide measures for certain learning outcomes of the physics major programs.

The use of all publicly available course materials \citep{CapstoneGithub} is encouraged. They are available as full courses (for Physics Capstone and Physics Symposium) but also organized by theme for instructors who would like to focus on one or two areas. All materials and activities are developed for a relatively small number of students, 10 to 15 per class. Some assignments and activities are easily scalable to larger programs or adaptable to other areas of STEM, while others might require careful adaptation.

\section{Conclusions}
\label{sec:conclusions}

We have implemented changes to an undergraduate physics curriculum, with a focal point being written and oral communication. We have presented the development of a writing in physics curriculum based on writing pedagogy, carried out by physics faculty. This includes threshold concepts from writing studies such as genre, audience and purpose, the CAT approach, and structured peer feedback. Our approach led to significant improvements in the written and oral communication skills of the students, evident from the increase in the number of physics students obtaining competitive research fellowships and the increased quality of the final papers, posters, and presentations in the last semester before graduation. Preliminary research on the effects of specific interventions has been carried out, which will be expanded and reported on in the future. The course materials for all three courses that are part of the physics writing curriculum are publicly available \citep{CapstoneGithub}, and the use and further development of these materials by the community are encouraged.

\vspace{0.2in}

\begin{acknowledgments}

We thank Phillip Troutman for providing ideas for peer feedback activities in the Physics Capstone and Symposium courses, and Gordon Mantler for his support to push our writing in physics effort forward. We also thank members of the Writing in STEM group at our university, in particular Kylie Quave and Tiffany Sikorski, for the great discussions that have informed our thinking.

\end{acknowledgments}

\bibliographystyle{apsrev4-1}
\bibliography{WritingInPhysics}

\end{document}